\newcommand{\be}{\begin{equation}}
\newcommand{\ee}{\end{equation}}
\newcommand{\bea}{\begin{eqnarray}}
\newcommand{\eea}{\end{eqnarray}}

\documentstyle[12pt]{article}
\advance\textheight by \topskip
\begin{document}

\begin{flushright}
SU-ITP-98-46\\

hep-th/9807206\\
July 28, 1998
\end{flushright}
\vspace{.5cm}

\begin{center}
\baselineskip=16pt

{\Large\bf  Superconformal Actions in Killing  Gauge} \\

\

\vskip 1 cm

{\bf  Renata Kallosh}

\vskip 1cm

{\em Physics Department, Stanford
University, Stanford, CA 94305-4060, USA\\
kallosh@physics.stanford.edu
}

\end{center}

\vskip 1 cm
\centerline{\bf ABSTRACT}
\vspace{-0.3cm}
\begin{quote}

The classical superconformal actions of branes in adS superspaces  
have a closed
form depending on a matrix ${\cal M}^2$ quadratic in fermions, as found in
hep-th/9805217. One can  gauge-fix the local  $\kappa$-symmetry
using  the Killing spinors  of the brane in the bulk.  We show that  
in such
gauges the superconformal actions are simplified dramatically since ${\cal
M}^2=0$ in all cases.  The relation between {\bf classical} and {\it
gauge-fixed} actions for these theories reflects the relation  
between the full
{\bf superconformal} algebra and its {\it supersolvable}  subalgebra.

\end{quote}
\normalsize

\baselineskip=16pt

\newpage

The classical actions of supersymmetric branes are known in generic  
superspaces
of supergravity. The basic problem with the generic superspace is that the
vielbeins of the background are superfields, depending on 32 Grassmann
variables $\theta$. The actions are rather complicated to deal with.
Particularly interesting examples of the generic backgrounds of  
supergravity
are given by the near horizon superspaces \cite{KRR}. These  are exact
solutions of M-theory and IIB string theory in the superspace  
\cite{KR}. The
backgrounds are  purely bosonic with the  $adS_{p+2} \times  
S^{d-p-2}$ metric
and the form fields required for the unbroken supersymmetry.

The purpose of this note is to show that in such backgrounds the  
superconformal
actions of the branes can be gauge-fixed in the Killing gauge
\cite{KRR,K,eric}, which greatly  simplifies the theory. It is rather
impressive that the geometry of the consistent background seems to  
teach us how
to perform the quantization of the extended objects  which live in such
backgrounds!

The Killing gauge for superconformal brane actions has the following  
important
features:
\begin{itemize}
\item The classical actions have to be placed in AdS background  
described by
 the Killing gauge in superspace \cite{KRR}: the curved fermion-fermion
vielbein at vanishing fermionic coordinates is not a unit matrix as in the
standard Wess-Zumino gauge, but a matrix codifying the relation  
between the
maximal dimension space dependent Killing spinor  near the horizon and its
constant part.

\item  The local $\kappa$-symmetry has to be gauge-fixed using the  
projector
for the Killing spinor in the bulk  where the brane has only 1/2 of
supersymmetry unbroken \cite{K}.

\item Killing vector gauge fixing reparametrization symmetry  
(static, physical
gauge)  is  natural to  use in this package, but is not necessary.
\end{itemize}

The  coset superspace was constructed in \cite{KRR} using the  generic
superalgebra (for the near horizon superspace we are interested in a
superconformal algebra) of the form
\begin{eqnarray}
\left [B_A ,  B_B \right ] &=& f_{AB}^C B_C  \ , \nonumber \\
\left [F_{\alpha} ,   B_B \right ]  &=&  f_{\alpha B}^\gamma F_\gamma \ ,
\nonumber\\
\left\{ F_\alpha  , F_\beta \right\}&=&  f_{\alpha \beta}^C B_C \ ,
\label{algebra}\end{eqnarray}
where $B_A$ and $F_\alpha$ are the bosonic and fermionic
generators, and   $f$ are the structure constants. The nilpotent 1-form
differential operator ${\cal D}$ associated with the superalgebra
(\ref{algebra}) is constructed as follows:
\begin{equation}
{\cal D}_{cl} = d + L^\alpha F_\alpha + L^A B_A  \ , \qquad {\cal  
D}^2_{cl}=0 \
{}.
\end{equation}
 The vielbeins of such superspace are still superfields depending on 32
components of $\theta$, however as different from the general case, this
dependence can be  given in a closed form in terms of the matrix  
${\cal M}^2$,
quadratic in fermions \cite{KRR}. This matrix depends on the structure
constants of the superalgebra in F-F and F-B sectors.
\begin{equation}
\left({\cal M}^2\right)^\alpha_\beta =-\theta^\gamma f_{\gamma
A}^\alpha\theta^\delta f_{\delta \beta}^A \ .
\label{2d}\end{equation}
The vielbeins depend on particular functions of this matrix:
\begin{equation}
L^\alpha_{cl} =\left({\sinh {\cal M}  \over {\cal
M}}\right)^\alpha_\beta
(D\theta)^\beta_{cl} \ , \qquad
L^A_{cl} =L^A_0+2 \theta^\alpha  f_{\alpha \beta}^A
\left({\sinh^2 {\cal M}/2  \over {\cal
M}^2}\right)
^\beta_\gamma
(D\theta)^\gamma_{cl}\ .
\label{WZ2}\end{equation}
Here
\begin{equation}
(D \theta)^\alpha_{cl} \equiv  d\theta^\alpha +( L^A_0 B_A  \theta)  
^\alpha,
\end{equation}
and $L^A_0$ are the values of the bosonic part of the vielbeins at  
$\theta=0$.
The details of this construction as well as examples of the closed  
form of the
classical actions for the IIB string and D3 brane in the near horizon
superspace of the D3 brane can be found in \cite{MT1,KRR,MT2}.

In the flat superspace the matrix ${\cal M}^2$ vanishes. The reason  
is that in
this case the superalgebra is the superPoincar\'{e} algebra with the  
commutator
of two supersymmetries producing translation
\begin{equation}
\{Q, Q\} \sim P  \qquad \rightarrow \qquad f_{\delta \beta}^A\neq 0 \ .
\end{equation}
Supersymmetry commutes with translation
\begin{equation}
[Q, P ] \sim 0  \qquad \rightarrow \qquad f_{\gamma
A}^\alpha=0
\end{equation}
and therefore for the flat superspace
\begin{equation}
[Q \ , \{ Q \ , Q \} ] =0 \qquad \Longrightarrow \qquad \left({\cal
M}^2_{flat}\right)^\alpha_\beta =-\theta^\gamma f_{\gamma
A}^\alpha\theta^\delta f_{\delta \beta}^A =0 \ .
\end{equation}

The basic reason why in case of superconformal algebra ${\cal M}\neq  
0$ is the
following. The commutator of two supersymmetries is producing  
translation as
well as Lorentz transformation
\begin{equation}
\{Q, Q\} \sim P + M \qquad \rightarrow \qquad f_{\delta \beta}^A\neq 0 \ .
\end{equation}
Here supersymmetry does not commute with translation as well as with the
Lorentz transformation:
\begin{equation}
[ Q, P ] \sim Q \ ,  \qquad  [ Q, M ] \sim Q \qquad \rightarrow \qquad
f_{\gamma
A}^\alpha \neq 0 \ .
\end{equation}
Therefore for the near horizon superspace
\begin{equation}
[Q \ , \{ Q \ , Q \} ] \neq 0 \qquad \Longrightarrow \qquad \left({\cal
M}^2_{nearhor}\right)^\alpha_\beta =-\theta^\gamma f_{\gamma
A}^\alpha\theta^\delta f_{\delta \beta}^A \neq 0 \ .
\end{equation}

The natural question arises whether it is possible in case of the  
near horizon
superspace to gauge-fix the local $\kappa$-symmetry and reparametrization
symmetry to get rid of the matrix ${\cal M}^2$ altogether. The answer is
positive, universal and rather simple.\footnote{Recently  the  gauge-fixed
supermembrane action was constructed in \cite{Fre} on the basis of  
{\it Ssolv}
part of the superconformal algebra $Osp(8|4)$.
One  could  also start  with the classical action in the near horizon
superspace  and perform the gauge-fixing in the Killing gauge, which  
in this
case results
in vanishing of the matrix ${\cal M}^2$. }  Superconformal symmetry  
has two
types of supersymmetries, the usual supersymmetry $Q_{+{1\over 2}}$  
and the
special supersymmetry $S_{-{1\over 2}}$. Under dilatation $Q$ has a  
positive
charge $+{1\over 2}$ and the special supersymmetry has a negative charge
$-{1\over 2}$:
\begin{equation}
[ D, Q_{+{1\over 2}} ] = {1\over 2} Q_{+{1\over 2}} \ ,  \qquad  [ D,
S_{-{1\over 2}} ] = -{1\over 2} S_{-{1\over 2}} \ .
\end{equation}
If one would consider only the subalgebra of the superconformal  
algebra which
includes positively charged under dilatation supersymmetry  
$Q_{+{1\over 2}}$
and excludes the negatively charged $S_{-{1\over 2}}$ one would  
arrive at the
following simple picture (we will use the spit of the spinorial  
index $\alpha$
into $\alpha_+$ and $\alpha_-$)
\begin{equation}
\{Q_{+{1\over 2}} , Q_{+{1\over 2}} \} =  f ^{\; \; + 1 A}_{+{1\over 2}\;
+{1\over 2}}
B_{+1A} \qquad \rightarrow \qquad (f_{\delta_+ \beta_+}^{+1A})_{sub}  
\neq 0
\label{supersymmetry} \ .
\end{equation}
Here $B_{+1A}$ are some bosonic operators with dilatation charge $+1$
\begin{equation}
[ D, B_{+1A} ] =   B_{+1A} \ .
\end{equation}
This leads to
\begin{equation}
[ Q_{+{1\over 2}} , B_{+1A}   ] =0     \qquad \rightarrow \qquad  
(f_{\gamma_+
A}^\alpha)_{sub} = 0 \ ,
\end{equation}
since the algebra does not contain any odd generators with  
dilatation charge
$+{3\over 2}$.
Note that the usual supersymmetry $Q_{+{1\over 2}}$ is the one which  
is present
in the brane in the bulk. Only near the horizon where the conformal  
symmetry
appears the special supersymmetry also appears, in particular in the  
commutator
of usual supersymmetry with the special conformal symmetry. It is clear
therefore that if one would constrain the Grassmann coordinates of the
superspace by the same requirement as the one which gives  ${1\over 2}$ of
unbroken supersymmetry in the bulk, one would exclude special  
supersymmetry
from the superconformal algebra. The remaining subalgebra will be  
such that the
structure constants of the subalgebra will give a vanishing matrix ${\cal
M}^2$.
\begin{equation}
[Q_{+{1\over 2}}  \ , \{ Q_{+{1\over 2}}  \ , Q_{+{1\over 2}}  \} ]  
=0 \qquad
\Longrightarrow \qquad \left [\left({\cal M}^{ 2
}_{nearhor}\right)^{\alpha_+}_{\beta_+}\right ]_{g.f.} =-\theta^{\gamma_+}
(f_{\gamma_+
A}^\alpha)_{sub} \; \theta^{\delta_+} (f_{\delta_+ \beta_+}^A)_{sub} = 0.
\end{equation}
This is the crucial point for  simplification of the fermionic part of the
superconformal gauge-fixed brane actions.

We will proceed here with the details of the general construction of the
gauge-fixed superconformal actions starting with the classical  
actions in the
near horizon superspace background. This outlines the general  
strategy for the
supersymmetrization of the conformal actions of the branes constructed in
\cite{conf}. The cases covered in bosonic case include M2, M5, D3 branes,
self-dual 6d string, magnetic string and electric black holes in 5d and
electromagnetic black holes in 4d. The geometry is that of  
$adS_{p+2}\times
S^{d-p-2}$ with the form field. Our strategy for supersymmetrization  
will apply
to all these cases.

Let us clarify  some important properties of the Killing spinors near the
horizon and in the bulk.
The bosonic coordinates $X$ include $p+1$ directions of the brane, $x^m,
m=0,1\dots , p$, the radial coordinate $\phi$ defining the bulk of the
$adS_{p+2}$ space and the angles $\eta_i$ of the sphere. As in  
bosonic case
\cite{conf} we will work in the Killing adapted frame where the  
metric of the
brane is independent on the Killing directions $x^m$:
\begin{equation}
ds^2=\phi ^2\,dx^m\,\eta_{mn}\,dx^n+(wR)^2 \left( \frac{ d\phi }{\phi }
\right)^2 +R^2 d^2\Omega \, ,
\label{induced}\end{equation}
where $d\Omega^2$ is the $SO(d-p-1)$-invariant metric on the unit
$(d-p-2)$-sphere with angles $\eta_i$. The constant $R$ is the radius of
curvature of the
$S^{d-p-2}$ factor, whereas the $adS$ radius of curvature is $wR$, where
$w = (p+1)/ (d-p-3)$.
It is also sometimes useful  to change the coordinates so that  $\phi=
\left({r\over R}\right)^{1\over \omega}$
\begin{equation}\label{cart}
ds^2=\left({r\over R}\right)^{2\over \omega}
\,dx^m\,\eta_{mn}\,dx^n+\left({R\over r}\right)^2  
dx^{m'}\delta_{m'n'} dx^{n'}
\ .
\end{equation}
Now the radial coordinate of the $adS_{p+2}$ together with angles of  
the sphere
 form an Euclidean part of the metric and we can use a cartesian
coordinate system.

The near horizon metric admits a maximal size Killing spinor
\begin{equation}
\epsilon^{\alpha} (x, \phi, \eta))_{Kill}^{Hor} = e^{\alpha} _{\underline
{\alpha}}(x, \phi, \eta)
\epsilon_{\rm const}^{\underline {\alpha}} \ ,
\end{equation}
where $\epsilon_{\rm const}^{\underline {\alpha}}$ is a constant  
unconstrained
spinor. The curved fermionic-fermionic vielbein of the near-horizon  
superspace
at $\theta=0$ is given by $e^{\alpha} _{\underline {\alpha}}(x,  
\phi, \eta)
$  for all cases of interest. It has the following structure \cite{LPR}
\begin{equation}
e^{\alpha} _{\underline {\alpha}}(x, \phi, \eta)= (f (\phi,
\eta_i))^\alpha_\beta
\left [ 1 -{1\over 2}  x^m  \Gamma_m  \Gamma_r (1- \tilde \Gamma)\right
]^\beta_{\underline {\alpha}} \ .
\label{ferm}\end{equation}

Away from the horizon the brane configuration admits only 1/2 size Killing
spinor which in the metric (\ref{induced}) has the form
\begin{equation}
\epsilon_{Kill}^{Bulk}( \phi, \eta) = (g_{tt})^{1/4}f(  
\eta)\epsilon_0 = \sqrt
{ \phi }f( \eta) \;  \epsilon_0 \ .
\end{equation}
The Killing spinor in the bulk has a simple form: it is proportional to a
quartic root from the time-time components of the metric times a  
matrix $f$
depending on angles times the constant spinor $\epsilon_0$.
Killing spinor in the bulk is satisfying some  algebraic constraint  
since it
describes 1/2 BPS state:
\begin{equation}
 (1- \tilde \Gamma) \epsilon_0   =0\ ,  \qquad  (1- \tilde \Gamma)
\epsilon_{Kill}^{Bulk}( \phi, \eta) =0 \ .
\end{equation}
It is easy to see that to project the Killing spinor in the bulk from the
Killing spinor at the horizon one has to impose the constraint
\begin{equation}
 (1- \tilde \Gamma)^\beta _{\underline {\alpha}}  \epsilon_{const}  
^{\underline
{\alpha}}    =0 \ .
\label{gauge}\end{equation}
This constraint simultaneously with projecting half of the spinor  
also removes
the dependence on Killing directions of the metric $x^m$ from the Killing
spinor  near the horizon. We define
\begin{equation}
{\cal P}_{\pm} \epsilon \equiv  {1\over 2} (1\pm  \tilde \Gamma) \epsilon
\equiv    \epsilon^{\pm} \ .
\end{equation}
The original expression  for the near horizon Killing spinor can be  
rewritten
as follows
\begin{eqnarray}
\epsilon^{+} (x, \phi, \eta))_{Kill}^{Hor} &=& e^{+} _{+}(\phi, \eta)
\epsilon^{+}_{ const} + e^{+} _{-}(x, \phi, \eta)
\epsilon^{-}_{ const} \ ,\nonumber\\
 \nonumber\\
\epsilon^{-} (x, \phi, \eta))_{Kill}^{Hor} &=& \hskip 1 cm 0  \hskip  
1 cm  +
e^{-} _{-}(x, \phi, \eta)
\epsilon^{-}_{ const} \ .
\end{eqnarray}
A remarkable feature of the fermion-fermion vielbein (\ref{ferm}) is the
following: it is a triangle matrix with the lower left corner  
vanishing! It
follows that by choosing the algebraic constraint on the Killing  
spinor near
the horizon in the form (\ref{gauge})
we achieve two important goals. At $\epsilon _{const}^-     =0$
\begin{eqnarray}
\epsilon^{+} (x, \phi, \eta))_{Kill}^{Hor} = e^{+} _{+}(\phi, \eta)
\epsilon^{+}_{ const} \ ,  \qquad
\epsilon^{-} (x, \phi, \eta))_{Kill}^{Hor} = 0  \ .
\end{eqnarray}
First of all,  the constraint on the coordinate dependent Killing  
spinor is the
same as the one on the constant part, which means that $\epsilon^{-}  
(x, \phi,
\eta))_{Kill}^{Hor} = 0 $.
Secondly,  the remaining part of the Killing spinor,
$\epsilon^{+} ( \phi, \eta))_{Kill}^{Hor} = e^{+} _{+}(\phi, \eta)
\epsilon^{+}_{ const}$ does not depend on Killing directions of the  
metric $x$
anymore.

These two facts will be important for  quantization.  The first one  
will allow
to have simultaneously vanishing $\theta^-$ as well as $L^-$. The  
second one
provides the possibility to use the physical, static gauge for the branes
$x^m=\sigma^m$. If the classical action would depend on $x^m$ (without
derivatives)  in the static gauge we would end up by the quantized action
which depends on coordinates of the brane explicitly and not only  
through the
fields. This would make the static gauge not acceptable for quantization
if the choice of the  Killing gauge in the superspace has been made.
Fortunately in Killing spinor gauge for  $\kappa$-symmetry, the  
dependence on
$x^m$ without derivatives drops.

Our choice of the gauge in superspace is the Killing gauge \cite {KRR} as
opposite to  the standard Wess-Zumino gauge  in superspace. For this  
we have to
consider space-time dependent $\theta^\alpha$ (and space-time independent
$\theta^{\underline
{\alpha}}
$)
of the form
\begin{equation}
\theta^{\alpha}(X) = e^{\alpha} _{\underline {\alpha}}(X)
\theta^{\underline
{\alpha}} \ ,
\end{equation}
so that
\begin{equation}
(D\theta)^\alpha =e^\alpha_{\underline \alpha}(X) d \theta ^{\underline
\alpha}
+\left (  (d+ L^A_0 B_A)^{\alpha}_ \beta  e^\beta _{\underline
\alpha}(X) \right
)\theta ^{\underline \alpha} \ .
\end{equation}
The second term drops if we choose $\theta^{ \alpha}(X)
$ to
be the Killing spinors of the background, satisfying  the equation
\begin{equation}
 (d+ L^A_0 B_A)^{\alpha}_ \beta  \epsilon^\beta  (X)_{Kill}^{Hor} =0 \ .
\end{equation}
We get
\begin{equation}
\qquad
(D\theta)^\alpha =e^\alpha _{\underline \alpha}(X) d \theta ^{\underline
\alpha}  \ .
\end{equation}
The superspace structure which is a background for the classical  
superconformal
actions in Killing gauge in superspace becomes:
\begin{equation}
L^\alpha_{cl} =e^\alpha_{\underline \alpha} (X, \theta) d \theta  
^{\underline
\alpha} \ , \qquad L^A_{cl} =e^A_M (X, \theta=0) dX^M  + e^A_{\underline
\alpha}(X,
\theta)
  d \theta ^{\underline \alpha} \ ,
\label{super1}\end{equation}
where
\begin{equation}
e^\alpha_{\underline \alpha} (X, \theta) =\left({\sinh {\cal M}
\over {\cal
M}}\right)^\alpha_\beta
 e^\beta_{\underline
\alpha}(X) \ , \qquad
e^A_{\underline \alpha}(X, \theta)
=  \theta^\alpha  (X)  f_{\alpha
\beta}^A
\left({\sinh^2 {\cal M}/2  \over {\cal
M}^2}\right)
^\beta_\gamma
e^\gamma_{\underline \alpha}(X) \ ,
\end{equation}
where ${\cal M}^2$ now contains the space-time dependent spinors
$\theta^{\alpha}(X)$
\begin{equation}
({\cal M}^2)^\alpha_{\beta  \;  cl}= -\theta^\delta (X)  f^\alpha _{\delta
A}\theta^\gamma
(X)   f^A _{\gamma \beta} \ .
\label{M2}\end{equation}

All superconformal algebras can be split into $(Q, S)$ basis where  
$Q$ is an
ordinary supersymmetry and $S$ is a special supersymmetry. Under  
dilatations
$Q$ charges have weight $+{1\over 2}$ and $S$ charges have weight  
$-{1\over
2}$.
The important fact about the Killing spinors in the bulk and near  
the horizon
is the following. Only near the horizon the conformal symmetry  
appears, and
therefore the special conformal symmetry and special supersymmetry.  
In the bulk
we have only 1/2 of the Killing spinor and this 1/2 is associated with the
usual supersymmetry and not with the special supersymmetry.

To proceed with gauge fixing we impose the spinor Killing gauge on  
the brane
\cite{K,eric}:
we choose
\begin{equation}
\theta^-=0 \  \ .
\end{equation}
It was established in \cite{eric,K} that
the generators of $\kappa$-symmetry
\begin{equation}
\delta \theta = (1-\Gamma)  \kappa
\end{equation}
at $x^m=\sigma^m, x^{m'}=const, \theta=0$ reduce to the Killing projectors
${\cal P}_{-}\kappa$
\begin{equation}
\delta \theta| _{x^m=\sigma^m, \; x^{m'}=const, \; \theta=0} = (1-\tilde
\Gamma)  \kappa \ , \end{equation} and therefore the gauge
\begin{equation}
{\cal P}_{+}\kappa=0 \ , \qquad {\cal P}_{-} \theta \equiv \theta^-=0
\end{equation}
is acceptable. Therefore the Killing spinor and Killing vector gauge
\begin{equation}
\theta^-=0 \ ,  \qquad x^m=\sigma^m
\end{equation}
is a good gauge to use as it leads to the well defined quantum  
actions as it
was found in general in \cite{K} and demonstrated for M5 and M2 branes in
\cite{CKV, Fre}. If one is willing to use the advantage of the  
Killing spinor
gauge $\theta^-=0 $ ( ${\cal M}^2=0$ ) without being committed to  
the static
gauge $x^m=\sigma^m$ this is also possible. In particular for the GS  
type IIB
superstring in adS superspace, one may still want to use other gauges for
reparametrization symmetry: the light-cone one or the conformal one.  
One has to
verify in each case that the combination of the Killing spinor gauge for
$\kappa$-symmetry with some other gauge for fixing reparametrization  
symmetry
does lead to a well defined quantum action. If this condition is  
satisfied, the
simplification of the action follows from the simplification of the  
background
superspace after we choose $\theta^-=0 $.

In this gauge the superspace structure simplifies dramatically. As  
explained
before, due to the properties of the algebra, the matrix ${\cal  
M}^2$ vanishes
when this choice of the gauge for $\kappa$-symmetry is made. Due to the
triangle structure of the fermion-fermion vielbein, in this gauge also the
$L^-$component of the fermionic vielbein form vanishes and the  
$L^+$component
of the fermionic vielbein acquires a simple form linear in remaining  
fermions
\begin{equation}
L_{g.f.}^-=0 \ , \qquad L_{g.f.}^+=e^+_+(\phi, \eta)  d\theta^+ \ .
\end{equation}
The bosonic forms are also simple and include only $\theta$  
independent terms
and terms linear in $\theta$ and linear in $d\theta$ in the Killing vector
directions:
\begin{equation}
 L^{+1A}_{g.f.} =e^{+1A} _m (\phi, \eta, \theta=0) dx^m  +  \theta^+
e^+_+(\phi, \eta)   f ^{\; \; + 1 A}_{+{1\over 2}\;
+{1\over 2}}
e^+_{+}(\phi, \eta)
  d \theta ^{+} \ .
\label{super}\end{equation}
Here $f ^{\; \; + 1 A}_{+{1\over 2}\;
+{1\over 2}}
$ are the structure constants of the subalgebra of the  
superconformal algebra
defined by the two $+{1\over 2}$ dilatation charge supersymmetries in eq.
(\ref{supersymmetry}). The expression for the vielbeins in eq.  
(\ref{super})
can be simplified to
\begin{equation}
 L^{a}_{g.f.} =\phi ( \delta^a_m dx^m  +  \bar \theta^+ \Gamma^a
  d \theta ^{+}) \ .
\end{equation}
Here the dependence on angles $\eta$ drops from the bilinear  
combinations of
the Killing spinors and the dependence on $g_{tt}^{1/2} $ is in front.
In all  other  directions $L^{A'}_{g.f.}$ do not get any dependence on
$\theta^+$  since  in the Killing adapted frame these directions do  
not belong
to the non-vanishing r.h.s. of the anticommutator of the two  
supersymmetries
$Q_{+{1\over 2}}$
\begin{equation}
 L^{a'}_{g.f.} =e^{a'} _{m'} (\phi, \eta) dx^{m'}  \ .
\end{equation}

A comment about the connection of the quantization of the superconformal
actions developed here and  the {\it Ssolv} algebra approach of  
\cite{Fre} is
in order. A deep relation between solvable coordinates and near horizon
geometries has been already studied in \cite{CCAFFT}. It has been  
also  pointed
out in \cite{Fre} that a solvable superalgebra {\it Ssolv},
containing solvable ({\it solv}) algebra can be found inside  
$OSp(8|4)$ by a
suitable projection of the fermionic generators from $Q$ to ${\cal  
P}_+Q$. A
supersolvable subalgebra is the one for which after some finite number of
supercommutator operations all supercommutators vanish.
For our purpose of proving that
${\cal M}^2$
 of the classical action vanishes upon quantization it was important to
establish that
\begin{equation}
[Q_{+{1\over 2} }\ , \{ Q_{+{1\over 2}} \ , Q_{+{1\over 2} }\} ] =0 \ .
\end{equation}
This has allowed us to prove that
$
(f^\alpha_ {\gamma +  \; +1A} )_{sub} (f_{\delta_+ \beta_+}^{+1A})_{sub}=0
$
which is sufficient to prove that
$
({\cal M}^2)_{g.f.}=0.
$
In this respect we see that the tremendous simplification of the  
gauge-fixed
superspace  indeed happened by the reason that a triple supercommutator of
supersymmetries remaining after gauge-fixing vanishes. Thus the  
gauge-fixed
theory in Killing gauge in Killing adapted frame is related to the  
classical
one as the {\it Ssolv} subalgebra to the full superconformal algebra.

In conclusion, we have shown that the superspace background for the
superconformal actions of branes in adS spaces after gauge-fixing in  
Killing
gauge is described by very simple vielbeins in the Killing adapted frame:
\begin{equation}
L_{g.f.}^-=0 \ , \hskip 4.2 cm  L_{g.f.}^+=e^+_+(\phi, \eta)   
d\theta^+ \ ,
\end{equation}
\begin{equation}
 L^{a}_{g.f.} =\phi ( \delta^a_m dx^m  +  \bar \theta^+ \Gamma^a
  d \theta ^{+}) , \qquad  L^{a'}_{g.f.} =e^{a'} _{m'} (\phi, \eta)  
dx^{m'}  \
{}.
\end{equation}

In specific cases the final form of the gauge-fixed actions still  
has to be
worked out.

\vskip 1 cm
I am grateful to P. Claus, J. Rahmfeld and   A.~Rajaraman for the valuable
discussions.  This work  was
supported by the NSF grant PHY-9870115.

\end{document}